      [1999/12/01 v1.4c Il Nuovo Cimento]
\documentclass{cimento}
\usepackage{graphicx}  % got figures? uncomment this
\title{Pile-up correction for the Swift-XRT observations in WT mode}
\author{T.~Mineo\from{ins:1}\ETC, 
P.~Romano\from{ins:2}, V.~Mangano\from{ins:1}, A.~Moretti\from{ins:2}, G.~Cusumano\from{ins:1}, 
V.~La Parola\from{ins:1},  E.~Troja\from{ins:1}, S.~Campana\from{ins:2},
 G.~Chincarini\from{ins:2},  G.~Tagliaferri\from{ins:2}, 
M.~Capalbi\from{ins:3},  M.~Perri\from{ins:3},  P.~Giommi\from{ins:3},
 \atque
D.~Burrows\from{ins:4}}
\instlist{\inst{ins:1}INAF - Istituto di Astrofisica Spaziale e Fisica Cosmica di Palermo, Italy
\inst{ins:2}INAF - Osservatorio Astronomico di Brera, Italy
\inst{ins:3}ASI  - Science Data Center, Italy
\inst{ins:4}PSU  - Department of Astronomy \& Astrophysics, USA }
\PACSes{\PACSit{98.70.Rz}{$\gamma$-ray source; $\gamma$-ray bursts}
\PACSit{01.30.Cc}{Conference proceedings}}
\begin{document}

\maketitle

\begin{abstract}
The detector at the focal plane of the Swift X-ray Telescope (XRT) 
supports four readout modes,  automatically changed on board, to cover
the dynamical range of fluxes and rapid variability expected from GRB
afterglows. The Windowed Timing (WT) mode is used for sources with flux higher
than a few mCrab  and is obtained by compressing 
10 rows into a single row, and then
reading out only the central 200 columns of the CCD. Point sources with a
rate above  $\sim$300 c s$^{-1}$ produce severe pile-up in the central region of the
Point Spread Function.  This paper presents three  methods to correct the effects
of the pile-up in  WT mode. On ground calibration results and data from the
very  bright GRB 060124 are used to define and test these methods. 
\end{abstract}

%\section{Introduction}

Pile-up in Charge-Coupled Devices (CCD) is produced by the detection of two 
or more photons within the same region at the same gate time interval. 
The detector is unable to resolve  individual signals and a single event with the sum 
of the pulse heights is detected \cite{ref:dav}.  
The effect of pile-up is a lowering of the
source rate and a hardening  of the observed spectrum.  
The detection area for
Swift-XRT\cite{ref:bur} CCD in WT mode is a 7 $\times$ 1 pixel string; 
the charge produced by an X-ray photon is 
spread over this region with several shapes. A classification of
these shapes is performed associating  a grade number to each of them: low grades
correspond to confined charge distributions and high grades to extended 
distributions. A further effect of the pile-up is to change the event grade
distribution with respect to the one expected from a faint source: 
pile-up events with low
grade can be recorded as a single higher grade event producing a
change in the grade distribution. 

Swift-BAT triggered GRB 060124 on a precursor  
and the XRT started observing it in WT mode;  during the first sequence
the count rate of the burst was high enough  to cause
pile-up in the data\cite{ref:rom}. 
Ground testing and calibration of the XRT integrated system carried out 
at the Panter laboratory showed that WT mode data are affected by pile-up
for point source intensities $>$250 c s$^{-1}$.

Three different methods have been developed to check for the presence of
pile-up and estimate the inner PSF region to be excluded. 
To study the methods for correcting the pile-up effects 5 rate intervals
during which the observed source count rate was $<$ 100,
100--200, 200--300, 300--400, and $>$ 400 c s$^{-1}$ were selected.

The first method is based
on analysis of the radial intensity profile. The region affected by pile-up is
determined comparing the detected Point Spread Function with the one obtained
from on-ground calibration. 
The region to be excluded from the analysis is the one that
presents a number of counts lower than the expected one (see Fig.1 left panel).

The second method  is based on the effects  that pile-up generate on the spectral
distribution. 
The spatial region affected by pile-up is determined comparing the photon
index detected in regions with a central region of increasing size excluded
from the accumulation. The source is no longer piled-up when the spectral index does not show
variations (see Fig.1 central panel). 

The last method  is based on the effects  generated on the grade
distribution. 
A good approximation of the expected grade distribution with no pile-up is
obtained when the deficit at grade zero, associated with an increase at higher
grades, becomes negligible (see Fig.1 right panel).

All the three different methods  give 
equivalent results   
for the size of the central region that must be excluded
(4 pixels at rates $>$ 400 c s$^{-1}$ for the specific case of GRB 060124).

\begin{figure}[t]
\hspace{-0.5truecm}
\includegraphics[height=4.2cm,width=5cm,angle=0]{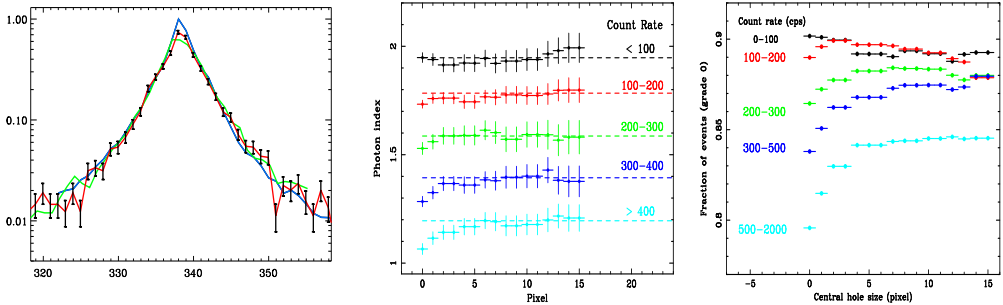} 
\hspace{0.1truecm} 
\includegraphics[height=4cm,width=4cm,origin=c,angle=-90]{mineot_fig1b.ps}
\hspace{0.1truecm} 
\includegraphics[height=4cm,width=4cm,origin=c,angle=-90]{mineot_fig1c.ps}
\caption{{\bf Left panel:} detected radial intensity profile
(points+connecting line for the count histogram  and light grey
 line for the relative fitting model) compared with the
expected one (dark grey line) \cite{ref:mor}. {\bf Central panel:} photon
indices detected at different count rates in regions with a central hole of
increasing size.  
{\bf Right panel:}  the grade
distribution detected at different count rates in regions with a central hole
of increasing size.
\label{fig:1}}
\end{figure}

\acknowledgments
This work is supported by the COFIN MIUR grant prot. number 2005025417, and
at Penn State by NASA contract NAS5-00136.


\begin{thebibliography}{0}
\bibitem{ref:dav} \BY{Davis J.}
  \IN{ApJ}{562}{2001}{575}.
\bibitem{ref:bur} \BY{Burrows~D. et al.}
  \IN{Space Science Rew.}{120}{2005}{165}.
\bibitem{ref:rom} \BY{Romano~P. et al.}
  \IN{A\&A}{456}{2006}{917}.
\bibitem{ref:mor} \BY{Moretti~A. et al.}
 \IN{Proceedings of the SPIE}{5898}{2005}{360}.
 \end{thebibliography}
\end{document}